# A Systematic Review on Interactive Virtual Reality Laboratory

1st Fozlur Rahman
*Department of Computer Science and Engineering*
*American International University-Bangladesh*
Dhaka, Bangladesh
shahariarahmed834@gmail.com

2nd Marium Sana Mim
*Department of Computer Science and Engineering*
*American International University-Bangladesh*
Dhaka, Bangladesh
mariumsana55@gmail.com

3rd Feekra Baset Baishakhi
*Department of Computer Science and Engineering*
*American International University-Bangladesh*
Dhaka, Bangladesh
feekrabaset@gmail.com

4th Mahmudul Hasan
*Department of Computer Science and Engineering*
*American International University-Bangladesh*
Dhaka, Bangladesh
shawkatkhanztuhin0022@gmail.com

5th Md. Kishor Morol
*Department of Computer Science and Engineering*
*American International University-Bangladesh*
Dhaka, Bangladesh
kishor@aiub.edu

*Abstract*—**Virtual Reality (VR) has became a significant element of education throughout the years. To understand the quality and advantages of these techniques, it's important to understand how they were developed and evaluated. Since COVID-19, the education system has drastically changed a lot. It has shifted from being in a classroom with a whiteboard and projectors to having your own room in front of your laptop in a virtual meeting. In this respect, virtual reality in the laboratory or Virtual Laboratory is the main focus of this research, which is intended to comprehend the work done in quality education from a distance using VR. As per the findings of the study, adopting virtual reality in education can help students learn more effectively and also help them increase perspective, enthusiasm, and knowledge of complex notions by offering them with an interactive experience in which they can engage and learn more effectively. This highlights the importance for a significant expansion of VR use in learning, the majority of which employ scientific comparison approaches to compare students who use VR to those who use the traditional method for learning.**

## I. Introduction

Technology is an ever-developing variant, there is no probable finishing line of ever-so innovations in this field. For years thousands of developments have amazed us. The use of technology in education has been really popular in recent years. ICT has contributed to activities in the educational sphere, with the help of technology interaction with a virtual environment is possible now with being a part of the environment in real-time [8]. The sector of Virtual Reality is also one of the major developments. Virtual reality has many potential extremities to uncover and so does augmented reality. Although augmented has to stay in a fixed place, virtual is spread out, but both have their unique advantages.

Students are more engaged in learning when they experience something in real life which can help them enhance and solidify their knowledge [1]. Virtual Reality can help students experience real-life environments and allow students to learn more visually, which is a more effective learning method. Virtual Reality has the potential to alter students' perceptions of learning [2]. Students can study visually while interacting with a specific setting in VR. VR helps them to develop and reinforce their knowledge. Virtual Reality allows students to access instructional materials from anywhere using contemporary gadgets [3]. With virtual reality extending to places where real life fails, Virtual Reality is a platform where people can be in their own comfort zone and still be able to access other dimensions per se. Virtual Reality has been making many advancements since the beginning, and although it takes a lot of input, the results are worth it. Basically, VR expands the level of creativity in a virtual platform, enabling people to run things according to their ways in a simulated dimension. Virtual Reality is slowly making more development towards perfecting its voids. As VR is still lacking in this area, the progress might seem very slow, all of it, though, has an influence.

Students sometimes face lots of issues in traditional classes while doing risky laboratory experiments, but we may easily overcome these obstacles with the aid of Virtual Laboratories [1]. With the help of the Virtual Laboratory, we don't have to worry about budgetary issues or lack of facilities in laboratories as the majority of schools and universities are unable to create them. Virtual Reality in Laboratories have been described as one of the best teaching methods that may keep students engaged during learning sessions. By using virtual reality, we can reduce the expense of laboratory experiments. But the main goal of Virtual Laboratory is to give users an experience of a 3D environment that depicts actual or imagined information and allows users to interact with them in real-time. In other words, it allows students to experience realistic laboratories in a virtual world where they can easily conduct experiments to solve issues in a manner similar to a traditional physical laboratory [7].

Although Virtual Reality or VR-based technology shows great potential for Student Learning and Laboratory settings. But we are all aware Virtual Reality is still a developing technology and with only a few VR applications created in the education sector [4] [5] and VR has sparked widespread interest. So, in the case of the laboratory settings, virtual reality shows lots of potential, but there are also a lot of variables, such as the cost of the system design, as well as its robustness, usability, and maintainability becomes a significant task [6].As a result, to optimize Virtual Reality's contributions in laboratory settings, some instructional methods must be established.

Our paper focuses on a specific part of virtual reality, which is, creating virtual platforms where students can interact for their lab performances be it chemistry, physics, or biology. In addition, the paper will discuss some relevant works in the realm of virtual learning systems. The laboratory experiments and evaluation of the results are also included in the study paper from the developer's perspective. Although we are going to be emphasizing more on Virtual Laboratory, hopefully, our findings and methods can also be used for the other science subjects. We will also be going through technological and pedagogical requirements, such as different objects, different OS systems, different resources, Oriented Systems, and how VR methods can improve in the development of these processes, and how we can update the developed system in the future [5].

The following is the paper's structure: In Section II, we talk about previous literature and reviews, in section III, we talk about materials and methods, in section IV, we talk about the study questions and discussion, and in section V, we talk about the conclusion

## II. Literature Review

In this section, we'll go through all of the prior work on Virtual Laboratories that has been done.

Several studies on Virtual Reality have been published in educational sections. The most recent was done by Jensen and Konradsen (2018) [9], who conducted a thorough analysis on the use of virtual reality. Their research provides a thorough examination of various learning outcomes as well as the user experience in Virtual Reality. Jensen and Konradsen (2018) [9] conducted research that demonstrates all 21 distinct qualitative and quantitative paper analyses that bring out both student learning and experience. When compared to less immersive technologies, the effectiveness of HMDs in the learning of analytic and operative skills was shown to be restricted in the paper. But due to the poor quality of research included, which made it really difficult to find any clear outcome regarding the educational usefulness of Virtual Reality [9].

A virtual based experiment was conducted in Indonesia at the University of Pendidikan Indonesia as a laboratory experiment-based learning proved to be highly efficient in teaching chemistry to students due to a lack of laboratories and equipment in some schools in Indonesia. One group was asked to answer 20 multiple-choice questions in the virtual-based experiment. Students in grade XII science at Bulukumba Regency's senior high schools in South Sulawesi Province, Indonesia, were the targeted group. The experiment's major objective was to assess the efficacy of Virtual Reality in the chemistry laboratory. 10 male and 20 female students were chosen from the group for the experiment. The average results of the students' experiment were 42.5 and 81.33, respectively, after the experiment was completed. There were 25 students that received scores of 75 or above, resulting in a completion rate of 83.33 percent. Finally, all of the data analysis results satisfied the above-mentioned efficacy requirements. To put it another way, the virtual laboratory is put to good use in a laboratory setting [10].

According to the current research In Nigeria, some students were tasked with doing volumetric analysis and responding to questions. The collected data were analyzed, and the findings reveal that there is a significant difference among both physical and virtual chemistry laboratory participants. The students who practiced in virtual chemistry laboratories fared considerably better than those who just practiced in real chemistry laboratories. It showed that three-dimensional interactive nature is very effective and efficient and also familiar with the actual procedures at work. Virtual chemistry laboratories solve the issue of insufficiency of facilities. As a result, practicing in a virtual chemistry laboratory is more supportive and beneficial in terms of improving students' comprehension and success [11].

A study was conducted at University College Copenhagen in Denmark. The main goal of the study was to see if Virtual Reality can be used as a simulation setting in practical education. The main objective was to improve new techniques to inspire students and improve their learning. The study was conducted on 78 students where various assessed educational elements of Virtual Laboratories were used in a 2-week course. Students perceived the ability to apply experimental simulation as well as the specific scenarios they were studying. The Virtual Laboratories appear to assist and help students apply theory to cellular functions, and also extensive laboratory techniques and equipment methods that should be practiced and visualized. As per the results of the study, the use of virtual reality in learning has the potential to enhance students' study engagement and interest. Virtual Reality is a really useful addition. This setting helps students move from traditional study methods to more advanced study methods [12].

A system was introduced named VR2E2C for Virtual Reality Remote Education in Chemistry. The main goal of the system design was to improve intellectual curiosity among public school and high school students by offering an enhanced education in conceptual and applied chemistry. In the system, there is a laboratory alternation option where disabled students can easily perform chemical experiments that have restricted mobility for regular use or experimentation. The program can be run in 3 ways. One option provides an intelligent robot controlled by the user to complete the task, which can perform the task by itself. The result can also be viewed by users in real-time. The system demonstrates how the newest Virtual Reality technology can help traditional chemistry teaching by providing remote teaching with safe experimental outcomes [14].

Some researchers from Spain conducted a bibliometric analysis in which they used two bibliometric approaches: performance analysis and scientific mapping. Researchers surveyed the VR lab with specific years of data ranging from 2015 to 2021 in order to add more advanced investigations and examine how those subjects have evolved over time. They acquired bibliometric data from various digital libraries. [15].

In 2021, several researchers released an article addressing the inadequacy of involvement among distant learners and instructors in higher education. Students must be properly instructed in order to complete the laboratory activity. Students must have fundamental and advanced understanding in order to conduct experimental tasks. As described in COVID-19, people from all over the world are trapped in a distant place. Everything is done online these days, and the remote lab is one of them. So the researchers used the Transitional Distance Theory to improve communication and, by contrasting traditional laboratory platforms and remotely active platforms, a case study on testing machines was performed. When the platform was integrated into regular learning, different users finished work in less time and boosted scores by more than 200 percent [16].

Research was conducted about a multi modal virtual chemistry laboratory. Multi modal virtual chemistry laboratory actually provides the specifics of a chemical object's physical and chemical qualities. The 3D interaction interface allows the user to effortlessly engage with MMVL. In this research, there are some questionnaires filled out by two groups of students. One is an MMVL trained group, and the other is an untrained group. Primarily, they have identified the range of effects, equipment, and their activities, as well as correctly performed experiments in a practical situation. When evaluating the MMVL instructed group to the unskilled group, there is a significant difference in overall percentages. According to the findings, untrained students learn at a rate of 32.7 percent, whereas trained students learn at a rate of 83.5 percent. Experiments show that students who received MMVL training have higher levels of confidence in the practical area than those who did not [17].

An interactive data integration feature of the architecture is highlighted in a study released by several researchers on the design of a virtual laboratory setting, as well as a demonstration of its use in the practical biology sector. Researchers are placing more emphasis on the remote laboratory project, which began at the University of Berlin and aimed to develop a hardware system proposal framework as well as an accessible, versatile, and customized laboratory framework to aid scientists and researchers in their exploratory work. The researchers explain the VL framework, its capabilities, and how they may be structured to make full use of its potential [19].

In a very recent paper published by University of Kanjuruhan Malang, the researchers discuss the factual improvisations of the subjective matter "Virtual laboratory for primary school students." In scientific education, experimental activities are sometimes limited by a lack of tools, finances, settings, and availability. These constraints inhibit the accumulation of scientific evidence, theories, and perceptions. The focus of the thesis was to construct an immersive environment that could be used as a learning tool. When developing a virtual lab, the procedure was to: identify the problems, operate the system, develop, evaluate, run, and maintain the system, and conduct analysis. In primary school science lessons, this medium is dependable and straightforward to use. Observation and questionnaires were utilized as tools in this research. The study's participants were fourth-grade pupils. Expert validation and practitioner feedback revealed that this medium is viable and suitable for use in scientific education. For elementary school children, virtual laboratory learning tools allow them to visualize abstract notions of material or occurrences into more tangible conceptions. Students' interest in studying science may be sparked via engaging and tangible learning tools. [20]

Some researchers published a paper on the effectiveness of e-learning with virtual laboratories considering the pandemic situation of 2020. The purpose of this research is to see how effective a VR lab is in an educational setting. The study's goal is to look at how a virtual lab might improve students' learning abilities and their knowledge of ideas in Chennai schools. The research also seeks to determine if the virtual lab aids students in increasing their self-paced learning. Surveys and expert interviews were used as research approaches. The results of the survey indicated that a high percentage of students are familiar with virtual laboratories and appreciate them highly. According to the report, virtual laboratories should be used in schools to encourage pupils to think beyond the box [21].

### III. MATERIALS AND METHODS

Different phases were conducted to reflect the study according to the approach for systematically analyzing research. In this section, we'll go over the stages. [18]:

#### A. Research Questions

In order to acquire the essential data, research questions were devised with the goal of knowing the emphasis of the study that is being conducted. The study questions we've outlined are related to Virtual Reality.

TABLE I
RESEARCH QUESTIONS

| Research Questions |
| --- |
| How many studies on VR Laboratory have been performed throughout the years? |
| In which areas of knowledge do students' believe Virtual Reality in laboratories should be employed in educational Institutes? |
| What is the evaluation method of VR users' learning and what is the learning outcome? |
| Which technological equipment can be used in the development of an interactive virtual laboratory? |
| What varieties of design tools are available for development and use in the study? |
| What will be the architecture of the Virtual Reality Laboratory if the research was implemented? |
| What are the limitations in the use of VR in the Laboratory? |

### B. Review of the Objective

The Review of the Objective will be evaluated a set of rules. They are Audience, Interaction, Comparisons, Results, and Context.

- **Audience**: The Audience of the Research is Universities and Students of different educational institutes.
- **Interaction**: Making Virtual reality a learning Environment
- **Comparisons**: We really cannot compare technologies.
- **Results**s: Understanding of Virtual Reality experience, knowledge, and technology outcome.
- **Context**t: Evaluation of users' Virtual Reality experience.

### C. Database and Search keywords

A literature search in an online database is one of the initial stages in every research project. As a result, in November 2021, We used a scientific database to do our search. To find publications about Laboratory Education in virtual apps or games, we used a collection of search phrases. The search string or phrases such as "Virtual Reality", "VR", "Virtual Laboratory", "Laboratory Experiment", "Learning", " Virtual Learning" was created with the help of the Research Questions we created. The database was chosen using digital sources since they offer a powerful search engine, which is required for the study. Between 2016 and 2021, the search was done utilizing the following databases: ACM, IEEE, BU library, Springer, and ScienceDirect. The combination of search terms that have been used are:

- ("virtual reality" OR "VR") AND ("laboratory" OR "lab")
- ("virtual reality" OR "VR") AND (educational OR "learning")
- ("virtual reality" OR "VR") AND (" Virtual laboratory" OR "VR lab") AND (""educational" OR "learning")

Table II illustrates the digital library and the search string or phrase used to identify relevant articles and researches based on title, abstract, and author Keywords.

TABLE II
DATABASE SEARCH STRINGS

| Source | Search Phrase or String |
| --- | --- |
| ACM Digital Library | [[Title: "virtual reality"] OR [Title: "vr"]] AND [[Title: educational] OR [Title: "learning"]] AND [[Abstract: "virtual reality"] AND [[Keywords: "virtual reality"] OR [Keywords: "vr"]] AND [[Keywords: "laboratory"] OR [Keywords: "lab"]] AND [[Title: educational] OR [Title: "learning"]] |
| IEEE Xplore | ("Title":"virtual reality" OR " Title":"VR") AND ("Title":"laboratory" OR " Title":"lab") OR ("Abstract":"virtual reality" OR "Abstract":"VR") AND ("Abstract Keywords":educational OR "Abstract Keywords":"learning") AND ("Abstract Keywords":"laboratory" OR "Abstract Keywords":"lab") OR ("Author":"virtual reality" OR "Author":"VR") AND ("Author":educational OR "Author":"learning") AND ("Author":"laboratory" OR "Author":"lab") |
| Science Direct | ("virtual reality" OR "VR") AND ("laboratory" OR "lab") AND ("educational" OR "learning"). |
| BU library | keywords ("virtual reality" OR "VR") AND keywords ("laboratory" OR "lab") AND keywords ("Education" OR "Educational" OR "Learning") |
| Springer | 'Virtual Reality AND VR AND laboratory AND learning AND "Virtual Reality laboratory" AND (VR)' |

Table III displays the overall number of publications and papers connected to our issue based on search keywords employed in various digital liberties, which is a total of 235 articles.

TABLE III
SEARCH RESULT OF DIGITAL LIBRARIES

| Source | Search Phrase or String |
| --- | --- |
| ACM Digital Library | 2 |
| IEEE Xplore | 11 |
| Springer | 47. |
| Science Direct | 70 |
| BU library | 106 |

### D. Study Selection

These criteria were prepared for the operation of collecting articles related to the study's topic are displayed in table below.

### E. Quality Evaluation

The evaluation criteria for the reduced number of papers are discussed in this section. We took into account the following questions to judge the papers for their quality:

- Are the article's goals related to virtual reality labs?
- Are the methods accurately explained for reaching the study's actual goal?
- Are all the results appropriately elaborate for a good overview of the carried research work of the study?
- Is the study experimental or theoretical?

TABLE IV
ACCOUNTED AND UNACCOUNTED

| Criteria | Criteria Description |
|---|---|
| Accounted | <ul><li>Articles from journals and relevant conferences emphasize the topic.</li><li>Articles with contents being all in the English language</li><li>Articles consisting of information about Virtual Reality Laboratories in their research</li><li>Theoretical papers with no applied research</li><li>Articles with systematic reviews</li><li>Articles that cited the main paper chosen</li><li>Articles that included past researches done on Virtual Reality</li></ul> |
| Unaccounted | <ul><li>Articles prior 2016</li><li>Articles that were not about Virtual Reality</li><li>Articles that did not contain anything about Virtual reality in Education</li><li>Articles that had less information on Virtual Laboratories</li><li>Articles that excluded VR Labs</li><li>Articles with less resourceful data</li><li>Articles that are focused on Mixed Reality</li><li>Articles that contained duplicated database</li></ul> |

- Was the study able to give an insightful end result or conclusive theory?

The Articles that were evaluated on the basis of these questions were eventually narrowed down to articles selectively.

*F. Data Extraction*

The following stages were constructed in order to acquire research data:

**Stage 1**: The first stage of the article selection procedure involves assessing all 236 articles based on their titles, publication year, and language. In the first stage of the article selection process, 17 papers are eliminated.

- Taking out the research that was exactly the same.
- Non-English studies are being eliminated.
- Reviewing and confirming the publishing year from 2016 to 2021.

**Stage 2**: The second step focuses on rejecting any research papers that do not integrate Virtual Reality Laboratory or Virtual Reality in student learning, as well as any studies that are irrelevant. In the second phase, 129 papers were eliminated.

- The studies that centered on augmented reality were discarded.
- Discarding papers from independent research.
- Research articles that do not include a Virtual Reality laboratory or Virtual Reality in student learning were discarded.

**Stage 3**: The third stage entails a thorough examination of the selected papers, during which 58 articles were eliminated.

- Effectively removing study papers that merely have the phrase "VR Lab" on them, but not using them anyhow.

- Articles which does not concentrate on either "Education" or "laboratory Education" are discarded.
- Articles which does not concentrate on Virtual Learning are discarded.

## IV. DISCUSSION

This research has a lot of potential in terms of student learning, and the VR Laboratory has a lot of potential in the future. We begin by responding to research questions based on 32 papers obtained from different digital library studies' findings in this study of Virtual Reality in Education.

*A. RQ1: How many studies on VR Laboratory have been performed over the years?*

Out of many papers skimmed, there were 32 relevant papers that were published between 2016 and May 2021 that were relevant to the study topic. 9 studies were published in 2016 (28.125%), 3 were published in 2017 (9.375%), 7 studies were published in 2018 (21.875%), 4 studies were published in 2019 (12.50%), 5 studies were published in 2020 (15.625%), and 4 studies were published up to May 2021 (12.50%) [Figure 1].

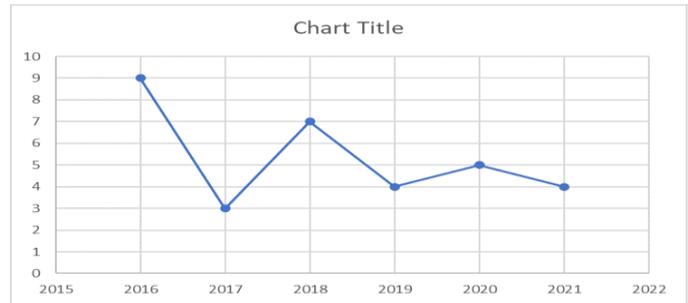

Fig. 1. Study on VR in education by year

*B. RQ2: In which areas of knowledge do students believe Virtual Reality in laboratories should be employed in educational institutions?*

In our research survey, we have received different responses from school, college, and university students. We have received a total of 110 responses [Figure 2] From university students, we have received 46 responses, which is 41.8% of the survey population. Among the 46 responses, 77.8% of students knew about virtual reality and 54.1% of the students used VR devices. 41.9% of students think that virtual reality can be employed in graphic-related courses. 40.3% of students think that virtual reality can be employed in EEE courses. 33.9% of students think that virtual reality can be employed in programming courses. 46.8% of students think that virtual reality can be employed in chemistry courses. 37.1% of students think that virtual reality can be employed in biology courses. 30.6% of students think that virtual reality can be employed in physics courses [Figure 5].

From college students, we have 31 responses. From the response, 72.5% of students know about virtual reality and 27.5% of students have used VR devices. 53.8% of students think that virtual reality can be employed in chemistry courses. 55.8% of students think that virtual reality can be employed in biology courses. 40.4% of students think that virtual reality can be employed in physics courses. 36.5% of students think that virtual reality can be employed in ICT courses [Figure 4]. From school students of the 8th to 10th standard, we have received 33 responses. 90.9% of students know about virtual reality and 16.1% of students have used VR devices. 54.5% of students think that virtual reality can be employed in chemistry courses. 30.3% of students think that virtual reality can be employed in biology courses. 45.5% of students think that virtual reality can be employed in physics courses [Figure 3].

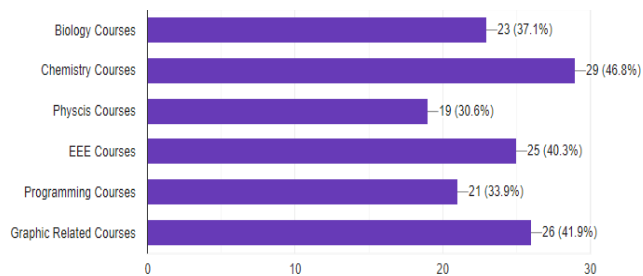

Fig. 5. Survey on University students

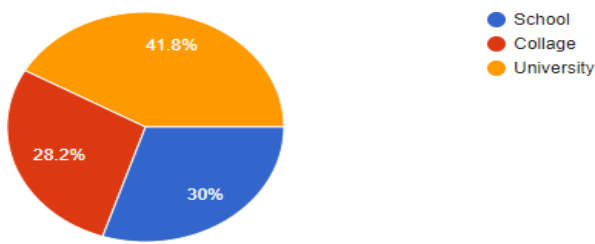

Fig. 2. Survey on different institutes

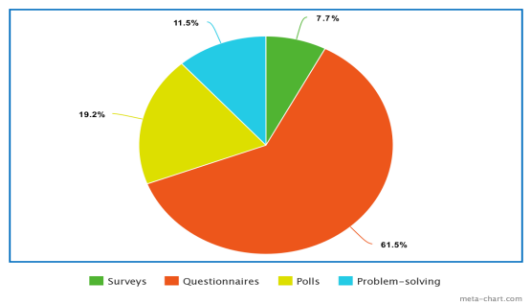

Fig. 6. Evaluation Methods

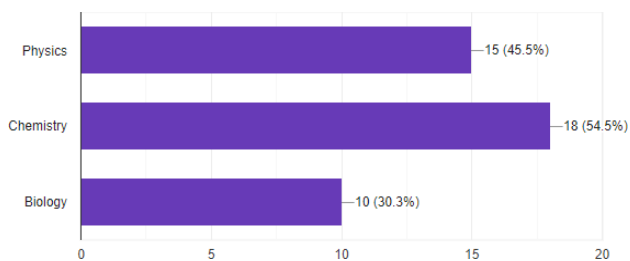

Fig. 3. Survey on School students

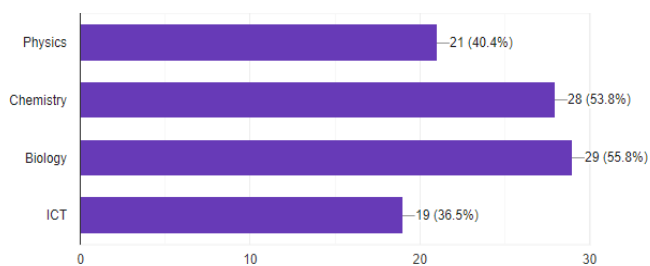

Fig. 4. Survey on Collage students

### C. RQ3: What is the evaluation method of VR users' learning and what is the learning outcome?

Questionnaires, polls, surveys, and problem solving techniques will be used to determine the evaluation technique of users' VR learning. Table V lists the evaluation procedures as well as all of the necessary evaluation papers. Questionnaires performed the majority of the evaluation, as seen in the table. From 2016 to 2021, 17 papers were assessed, and Figure 6 demonstrates that questionnaires with open and closed- ended questions account for 61.5% of the entire assessment technique. Polls, on the other hand, are the second-best way of analyzing VR users' experiences. According to the graph, polls were used for 19.2% of the evaluation. Another technique of evaluation is through surveys. After the polls, surveys show an 11.5% turnout. Students' thoughts on the learning experience were gathered through surveys. Finally, 7.7%, or the lowest evaluation, was done through problem-solving, which should not be so low, but according to the 17 papers examined for the research questions, only three articles were completed, one in 2017 and two in 2020. The problem is, it is unquestionably one of the greatest ways of appraisal, but not in this case. However, virtual reality is a relatively new technology, and while many papers have been published over the years, the majority of them lack problem-solving strategies, resulting in a 7.7% figure.

TABLE V
EVALUATION METHODS BY YEAR

| Evaluation methods | 2016 | 2017 | 2018 | 2019 | 2020 | 2021 |
|---|---|---|---|---|---|---|
| Surveys |  | [22] |  |  | [23] |  |
| Questionnaires | [24] | [25] [26] [27] | [28] [29] | [30] [31] [32] [33] [34] | [35] [36] [37] [38] | [39] |
| Polls | [24] | [25] [26] [27] | [28] [29] | [30] [31] [32] [33] [34] | [35] [36] [37] [38] | [39] |
| Problem-solving |  | [45] |  |  | [46] [47] |  |

*D. RQ4: Which technological equipment can be used in the development of an interactive virtual laboratory?*

TABLE VI
VR LABORATORY EQUIPMENTS

| Gears | 2016 | 2017 | 2018 | 2019 | 2020 | 2021 |
|---|---|---|---|---|---|---|
| HP REvarb | [24] [40] | [25] [43] | | | [38] | [48] |
| Pimax 8k | | | | [30] [32] | [23] | [37] |
| Oculus Quest | [49] | [26] [27] [42] | [29] | | [37] | |
| Gear VR | | [27] | | | [36] | |
| STEAM | | [27] | [44] | | [46] | |
| Valve Index | [49] | | | | [23] | [48] |

Table VI depicts the various technologies that can be employed in the building and research of virtual reality labs. From 2016 through 2021, the table covers the analyses conducted using various VR headsets. According to the table, 20%t of the work was done with HP REvarb, 16.7% with Pimax 8k, 20% with Oculus Quest, 8.3% with Gear VR, 12.5% with STEAM VR, and 12.5% with Valve Index. These devices were employed in the creation of the Virtual Reality Laboratory as well as research projects. From 2016 to 2021, many VR technologies were developed and various designs were applied using various VR devices and gears. The table provides thorough information on all of the gears utilized in the Evaluation method's 17 papers.

*E. RQ5: What varieties of design tools are available for development and use in the study?*

Table VII lists the tools required for designing a VR laboratory. Table VII is divided into four categories. The table below lists the software and programming languages that aid in the development of a virtual reality system. The table also lists the applications that are available on Virtual Laboratory, and the Others section lists all of the Virtual Reality applications that are related to either education or

TABLE VII
DESIGN TOOLS

| Programming language | C++, C JAVA, JavaScript, Python, Swift |
|---|---|
| Softwares | Unity, Amazon Sumerican Overview, UnrealEngine, Blender, 3ds Max, Maya, Oculus Medium, Matlab, AutoDesk3D |
| Application | Labster, Eonreality, Merck, VlabAcademy, IVF Laboratory VR, MEL VR |
| Others | InMindVR 2, TitanOfSpace, Anatomyou, KingTut VR, Unimersiv, TheBodyVR, 4Danatomy, ClassVR, DiscoveryVR, TiltBrush |

Learning. The majority of VR software development was done using C and JAVA, the majority of VR software development was done using C and Java. The review also reveals that UNITY3D is the most popular and user-friendly programming software. Although MAYA and BLENDER were the most popular software for object creation and texture design. The most specialized software for VR development is UNITY3D, Blender, and MAYA.

*F. RQ6: What will be the architecture of the Virtual Reality Laboratory if the research was implemented?*

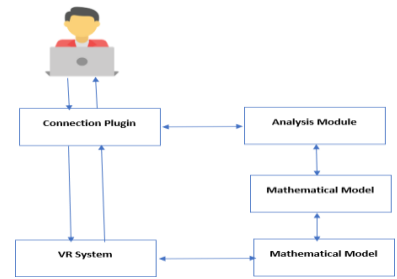

Fig. 7. Virtual Reality Laboratory architecture

The Architecture for the Virtual Reality Laboratory should be divided into six components.

- Connection Plugin
- Data Analysis Module
- Mathematical Models
- System Database
- VR System
- User Module

The Connection Plugin is the first stage of the architecture, and it can communicate with the User Module. The VR Connection Plugin has access to the user's actions when controlling the VR system. Because the Connection Plugin can control and interact with the User Module, it is linked to the VR System's Data Analysis Module. The Data Connection Module serves as a link between the user and the VR system. The Analysis Module will be linked to a database, which will assist the Analysis Module in distinguishing between users and keeping track of the users' entries for the VR System.

Both the Analysis Module and the Database will interact with mathematical models. The equations and data from the mathematical model will be used to build all of the equations and distinct experimental developments. The VR system will then serve as the primary virtual representation of various laboratory experiments. During this period, The virtual reality system serves as the user's environment. With the help of a connection plugin, users can experience a virtual environment. The system connection plugin will be triggered when the user runs. The Connection Plugin provides data through a data collection model that interacts with mathematical systems and databases and then displays the results on a VR display for the user [50] [Figure7].

*G. RQ7: What are the limitations on the use of VR in the laboratory?*

TABLE VIII
VR LABORATORY LIMITATIONS

| Limitations | 2016 | 2017 | 2018 | 2019 | 2020 | 2021 |
|---|---|---|---|---|---|---|
| Gear use problem | [40] | | | [31] | [35] [37] [48] | [36] |
| Understanding time | [40] | | | | [37] [35] | [36] |
| Gear Cost | [49] | | | | [35] [38] | |
| Nausea | | | | [44] | [37] | [36] |
| Connection Issues | | [22] [26] | | | | |
| Technical Issues | | | [21] | | | |

Table VIII shows the limitations of using the VR laboratory. The limitations include the gear use problem, which is 31.6%. Users face various gear usage problems. Another problem is the understanding time, which is 21.1%. The gear cost is one other big limitation, at 15.8%. Some users also face nausea, which happens to be 15.8%. Internet issues are 10.5%, and technical issues like software or hardware errors are 5.4%. The data was obtained from 17 papers included in the evaluation methods. After the paper review, these limitations were identified in the papers. There are also other limitations involved with using VR gear and VR lab equipment, such as communication issues, fewer users, not having the proper equipment or gears, and lack of budget.

## V. CONCLUSIONS

Based on studies published between 2016 and 2018, Virtual Reality (VR) has a lot of potential in laboratory education. We found 72 papers in which Virtual Reality was used in a variety of educational contexts, and we looked into how these approaches were created and assessed.

Between 2016 and 2021 VR matured, and it made significant contributions in the realm of education. Almost every study looked at used at least one educational method or theory, the most common being Collaborative Learning, Multimedia Learning, and Inquiry-based Learning. The development of VR has changed the way of learning. It has changed the learning experience of students. With 3D imaging and virtual world interaction and the use of interactive virtual objects, it is another major aspect discovered in the analysis of the papers. Virtual Reality (VR) has quickly become one of the most advanced technologies that may be utilized in a variety of educational settings, which corresponds to the findings of the New Media Consortium's most recent Horizon Report on the technology's development.

The VR laboratory is one of the most promising technological advancements in the educational setting. The paper discusses how VR labs can change student learning environments and experiences. The paper also focuses on designing a VR lab and describes how open-ended and close-ended questions are the most effective evaluation methods for VR laboratory design. The studies done from 2016 to 2021 show VR gears are really effective in terms of laboratory experimental learning. Among the VR gear used, HP REvarb is one of the most popular and used gear. Through the study, we get an idea of VR applications and programming languages to build a VR laboratory program. This study gives the architecture for VR laboratory design. There are different ways to develop a VR program, but one of the most popular programs is UNITY3D. With the help of UNITY3D, it is possible to follow the Architecture of the VR Lab and develop a VR system. Although there are lots of limitations that have been discussed in the paper, the study gives us a clear picture of VR laboratory design and development. For the evaluation of the student experience of VR interaction and use, different surveys have been conducted, which also formed a great focus of this research as well.

Although VR has developed significantly over the years, in the field of education it is still in the development phase and others' findings. Different VR lab applications are currently accessible for many smart devices, including iOS and Android. However, there are not many in educational contexts. The absence of adoption of a development strategy reflects the platforms' lack of unanimity.

Finally, the data analysis indicated that when it comes to educational multimedia resources in VR that are available for interaction, few VR techniques are keeping up with the constant evolution of technology and new requirements from users. In actuality, 60 to 80 percent of research is based on out-of-date resources such as photographs and phrases. This gives us an idea of how a VR lab is necessary in the field of education and for the change of learning experiences. However, we observed that in order to create and enhance Virtual Reality Laboratories in educational contexts, more efficient and trustworthy resources and data are necessary.

## VI. ACKNOWLEDGEMENT